\documentclass[showpacs,preprint]{revtex4}

\usepackage{graphicx}
\usepackage{dcolumn}
\usepackage{amsmath}
\usepackage{epsfig}

\newcommand{\ec}{{\cal E}}
\newcommand{\polc}{{\cal P}}
\newcommand{\mc}{{\cal M}}
\newcommand{\ecv}{\vec{\cal E}}
\newcommand{\bcv}{\vec{\cal B}}
\newcommand{\pcv}{\vec{\cal P}}
\newcommand{\mcv}{\vec{\cal M}}
\newcommand{\kv}{\vec{k}}
\newcommand{\rv}{\vec{r}}

\newcommand{\vnabla}{\vec{\nabla}}

\begin{document}
\author{D.~Bernard}
\address{Laboratoire de Physique Nucl\'eaire et des Hautes Energies,
 Ecole Polytechnique, IN2P3 \& CNRS, 91128 Palaiseau, France}
\author{F.~Moulin}
\address{Laboratoire de Physique, Ecole Normale Sup\'erieure, 94235,
 Cachan, France}
\author{F.~Amiranoff}
\address{Laboratoire pour l'Utilisation des Lasers Intenses, Unit\'e
 mixte N$^{o}$7605 du CNRS-CEA-Ecole Polytechnique-Universit\'e
 Paris 6, Ecole Polytechnique, 91128 Palaiseau, France}
\author{A.~Braun, J.P.~Chambaret, G.~Darpentigny, G.~Grillon, S.~Ranc}
\address{Laboratoire d'Optique Appliqu\'ee, CNRS \& ENSTA, Ecole
Polytechnique, 91128 Palaiseau, France}
\author{F.~Perrone}
\address{Dipartimento di Fisica -Pisa and INFN, Piazza Torricelli 2, 56100 Pisa, Italy}
\author{~}
\address{\em \large To appear in The European Physical Journal D}

\title{Search for Stimulated Photon-Photon Scattering in Vacuum}

\date{November 22, 1999}

\begin{abstract}
 We have searched for stimulated photon scattering in vacuum at a
 center of mass photon energy of 0.8~eV.
 The QED contribution to this process is equivalent to four wave
 mixing in vacuum.
 No evidence for $\gamma\gamma$ scattering was observed. 
 The corresponding upper limit of the cross section is
 $\sigma_{\text{Lim}}=1.5~10^{-48}\text{cm}^{2}$.
\end{abstract}



\maketitle

\section{Introduction} 

Photon-photon scattering does not occur in classical electrodynamics
because Maxwell's equations are linear in the fields.
In Quantum ElectroDynamics (QED), $\gamma\gamma$
elastic scattering is described in
lowest order by a fermion loop with four open photon lines (box diagram). 
At low energies ($\hbar\omega\ll m c^{2}$), the corresponding cross
section is $ \sigma_{\text{QED}} = (973/10125\pi)
\alpha^{2}r_{e}^{2}(\hbar\omega/m c^{2})^{6} $ where $\hbar\omega$ is
the center of mass system (cms) photon energy, $m$ is the electron
mass, $\alpha$ is the fine structure constant, and $r_{e}$ is the
classical radius of the electron\cite{detollis}.
This cross section is extremely small in the optical domain where high
brightness sources exist~:
$\sigma_{\text{QED}}[\text{cm}^{2}]=7.3~10^{-66}(\hbar\omega[\text{eV}])^{6}$.

QED is a well established theory. The derivation of
$\sigma_{\text{QED}}$ is not in question.
Furthermore, the contribution of the box diagram is needed to describe
the already observed Delbr\"uck scattering and the high precision
measurements of the electron and muon magnetic moment.

The interest here is in the search for possible non-QED new physics in low
energy $\gamma\gamma$ scattering.
A theoretical basis for this is possibly coming from composite photon
theory \cite{ind} or the exchange of an axion \cite{axion}.

A previous experiment using the head-on collision of two laser beams
at different wavelengths has obtained a limit cross section 
of $10^{-39} \text{cm}^{2}$ (at 95\% CL) \cite{z}.
Here we improve this result by nine orders of magnitude by
stimulating the reaction with a third beam
\cite{kroll,varfolomeev,dewar,grynberg}.
The QED contribution to this process is equivalent to four wave
mixing in vacuum.

\section{The choice of the configuration}

In elastic scattering, the values of the energies $e_{i}$ and wave
vectors $\vec{k}_{i}$ of the incoming photons satisfy the
energy-momentum conservation condition~: $e_{1}+e_{2}=e_{3}+e_{4}$,
$\vec{k}_{1}+\vec{k}_{2}=\vec{k}_{3}+\vec{k}_{4}$, where indices 1,2
denote the incoming photons, 3,4 the outgoing photons.

In simple (ie. non stimulated) elastic scattering,
the final state is determined by two parameters
 (eg. the Euler angles of the decay axis in the cms).
Here we stimulate the reaction by a third beam, with a wavelength
$\lambda_3$; this fixes one parameter.
The direction of beam \#3 must lie on the cone of allowed direction for
a scattered photon at $\lambda_3$.
This position on the cone then fixes the second parameter.
The signal is then searched for in the direction of
$\vec{k} _{4}=\vec{k} _{1}+\vec{k} _{2}-\vec{k} _{3}$.
We have chosen to use three IR beams
($\lambda_1=\lambda_2=800~\text{nm}$, $\lambda_3=1300~\text{nm}$), with
the signal expected in the visible
($\lambda_4=(2/\lambda_1-1/\lambda_3)^{-1}=577~\text{nm}$). 

In this configuration, the photons of the input beams that scatter in
the residual plasma or on the optics can be spatially and spectrally 
filtered out, and the signal can be easily detected.
For strong signal isolation, the wavelength of the signal is also
chosen to be far from the wavelengths of the harmonics of the input
beams, which are always present in a high intensity beam.
\begin{figure}[hbt]
\begin{center}
\includegraphics[width=0.46\linewidth]{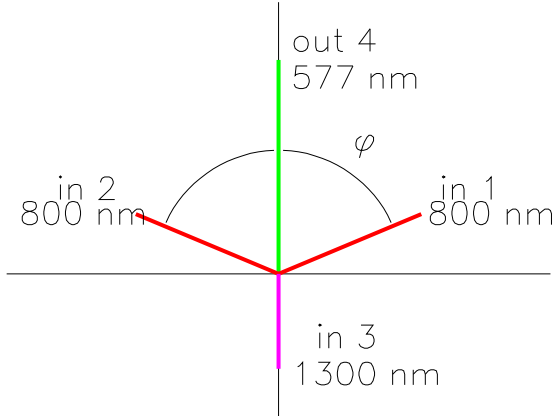}
\caption{Angular configuration of the stimulated experiment. The
 projection of the wave vectors $\vec{k}_{i}$ of the four beams on
 the pupil plane of the Bowen are shown.}
 \label{config}
\end{center}
 \end{figure}

The 3 beams are focused by a single optics made of a pair of spherical
mirrors (Bowen) with a coronal pupil of width 30~mm, and an equivalent
focal length of 100~mm.
The paraxial surface of the Bowen is a cone with a half angle of
33.6$^\circ$. A left-right symmetric configuration is chosen (fig.
\ref{config}) with the main beams at an angle $\varphi$ with respect
to the vertical direction; with $\cos(\varphi)=1-\lambda_1/\lambda_3$,
we have
$\varphi= 67.4^\circ$.
As the three beams are injected on that cone, the expected signal lies
also on the cone.
\begin{figure}[hbt]
\begin{center}
 \mbox{\epsfig{figure=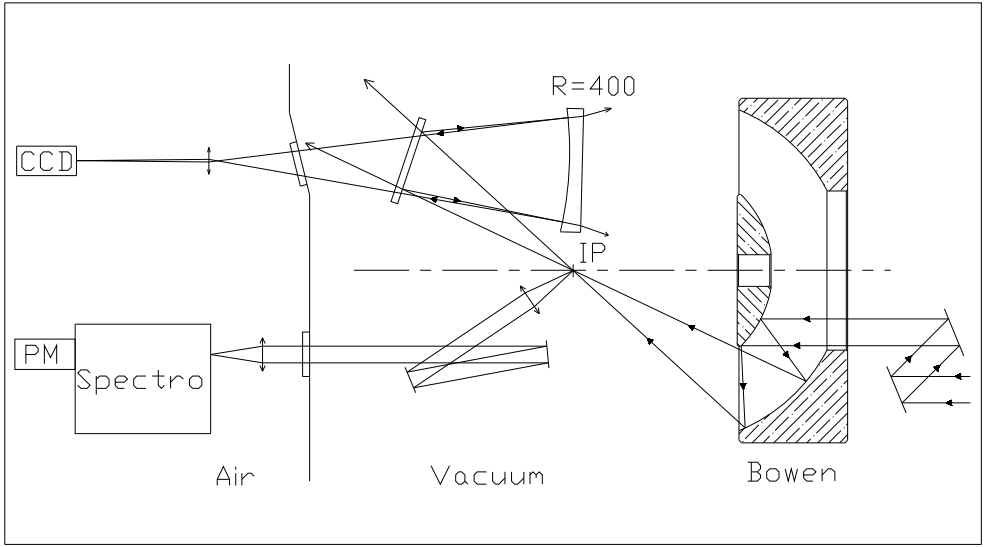,width=0.8\linewidth}}
\caption{Layout of the experimental apparatus. Only beam \#3 is shown.}
 \label{schemaexp}
 \end{center}
 \end{figure}

The characteristics of the beams are chosen so as to optimize their
overlap at the interaction point (IP).
The optimum is obtained for beams with the waist $w$ on the
order of the FWHM bunch length $c\tau$.

\section{Experimental Apparatus}

The two main beams (\#1 and \#2) at 805~nm are produced by a Ti:Sapphire
chirped pulse amplification (CPA)
 laser chain
with 3 amplifying stages\cite{chambare}.
Gaussian beams with 0.4~J of energy, 55~nm spectral width, and 40~fs FWHM
duration are delivered at 10~Hz.
A pair of tunable frustrated internal reflection attenuators allows 
the reduction of the intensity of the main beams without degrading
their optical quality or modifying their angle and temporal synchronisation.
Then, the two beams enter a vacuum chamber where they are expanded to a
diameter $\phi=30~$mm. The chirped pulses are temporally compressed to
40~fs, and are transported to the experimental chamber.

A fraction of the beam is collected after the second amplification
stage to pump the optical parametric amplifier (OPA) that produces the
beam at 1300~nm.

The 3 beams are injected into the Bowen with a diameter 
 $\phi$ of 30~mm.
A low intensity image of the focal spot of each beam is obtained
outside the chamber with a unit magnification by a combination of a
silica slide and of a non coated silica spherical mirror (fig. \ref{schemaexp}).
The image is enlarged by a microscope objective with a magnification
of 40 (20 for the 1300~nm beam for which the CCD sensitivity is low)
onto a CCD camera.

A set of dielectric mirrors with graded reflectivities at 800~nm
are used as filters down to an optical density of $D=3$, before and after the
objective.
Typical spot sizes at $1/e^{2}$ in intensity, $w$, of 4~$\mu$m (main
beams) and 6~$\mu$m (1300~nm beam) are obtained.

The expected signal photons are collected by a telescope with an $f$
number of 1.9, imaged onto the entrance slit of a spectrometer with a
transmission factor of 59~\% at 577~nm, and detected with a
photomultiplier (PMT) with an efficiency of 5~\%.
A BG38 filter further blocks the IR photons. The signal from the PMT
is 10~ns in duration at the foot and is digitized by a CAMAC ADC with
a gate of 25~ns.

\section{Experimental procedure}

The relative alignment and synchronisation procedures of the three
beams at the IP are dependent on each other due to the configuration
used.
They are performed in several steps.
First, the three beams are prealigned at IP on a single camera
located on the axis of the Bowen,
using a microscope objective with a large numerical aperture of 0.65, and their
waists are brought into a common plane perpendicular to that axis.
The position of the 3 spots in that plane is adjusted so as to
minimize their aberrations.
Then, a pre-synchronisation of the laser pulses is performed
with a precision of 25~ps with a fast diode and a 7~GHz oscilloscope.
The fine alignment is obtained by having each laser
punch the same hole in a 10~$\mu$m thick aluminum foil.
The synchronisation is then refined down to 100~fs by observing the
perturbation of the focal spot of a low energy beam after a plasma was created by a
high energy beam in a nitrogen gas jet with a pressure of about 0.3~bar.
At last, the fine synchronisation is performed by observing four wave
mixing ($\chi^{(3)}$) in the gas jet.

For this last step, the laser intensities are tuned just below plasma
threshold.
This method results in the spatial alignment and the synchronisation of
the beams with the upmost precision. Furthermore it maximizes the signal
in exactly the same configuration as that for the experiment in vacuum.
This point is detailed in the  following section.

\section{Four wave mixing and QED stimulated photon scattering}

Four wave mixing is a non linear process that appears in the
interaction of high intensity light beams in a medium.
The evolution of the fields in the medium are described by Maxwell's
equation in a non magnetic medium~:
\begin{equation}
\vnabla^{2} \ecv -
\frac{1}{c^2}\frac{\partial^{2}\ecv} {\partial t^{2}} = 
\frac{4\pi}{c^2}\frac{\partial^{2}\pcv} {\partial t^{2}}
\end{equation}
where the polarisation of the medium is developed as a function of
the field in the ``constitutive'' relations~:
\begin{equation}
\pcv(t) = 
  \underline{\chi}^{(1)}\ecv(t)
+ \underline{\underline{\chi}}^{(2)}\ecv^{2}(t)
+ \underline{\underline{\underline{\chi}}}^{(3)}\ecv^{3}(t)
+ ...
\end{equation}
Here we study the interaction of three incoming beams. We see that a
source term is present, that is proportional to $\ecv^{3}(t)$.
In particular, it contains a term proportional to
$\displaystyle e^{i(\omega_{4}t-\kv_{4}\cdot\rv)}$
with\footnote{In vacuum the four beams have the same phase velocity,
 so that the equation $\Delta \vec{k}=\kv_{1}+\kv_{2}-\kv_{3}-\kv_{4}=0$
 holds exactly. In a low pressure gas, the refractive indices at the
 four wavelengths are close enough so that $\Delta \vec{k} \approx 0$, i.e.
 more precisely $\Delta k\times w \ll 1$.}
 $\kv_{4}=\kv_{1}+\kv_{2}-\kv_{3}$ and $\omega_{4}=k_{4}c$.
A paraxial formulation for that component $\ecv_{4}$ along $\kv_{4}$,
in the slow varying wave approximation gives~:
\begin{equation}
\frac{\text{d}\ec_{04}} {\text{d} z} = -\frac{i \omega_{4}}{2c}
\chi^{(3)}\ec_{01} \ec_{02}\ec_{03}
~ ~ ~
\text{with}
~ ~ ~
\frac{\text{d}} {\text{d} z}= \frac{\partial} {\partial z}
+\frac{1}{c} \frac{\partial} {\partial t}
\label{danslegaz}
\end{equation}

Let's now turn to QED stimulated photon scattering in vacuum. The
insertion of the Euler-Heisenberg correction term \cite{lagrange} in Maxwell's
equations gives~:
\begin{equation}
\vnabla^{2}\ecv-\frac{1}{c^2}\frac{\partial^{2}\ecv}{\partial t^{2}} = 
\mu_{0}[\frac{\partial}{\partial t}\vnabla\wedge\mcv +
 \frac{\partial^{2}\pcv}{\partial t^{2}}-c^{2}\vnabla(\vnabla\cdot\pcv)],
\end{equation}
with
$\pcv=2a\left[2(\ecv^2-c^2\bcv^2)\ecv+7c^2(\ecv\cdot\bcv)\bcv\right]$,
$\mcv=2a\left[-2c^{2}(\ecv^{2}-c^{2}\bcv^{2})\bcv+7c^{2}(\ecv\cdot\bcv)\ecv\right]$,
and $a=\hbar e^4/(360\pi^2 m^4 c^7)$.
Under the same approximations as for 4 wave mixing in a medium, we get~:
\begin{equation}
\frac{\text{d}\ec_{04}} {\text{d} z} 
\vec{u}_{4}
= -\frac{i\mu_{0} \omega_{4}}{2}
[(c \polc_{0x}+ \mc_{0y})\vec{u}_{x}+
(c \polc_{0y}- \mc_{0x})\vec{u}_{y} ]
\end{equation}

\begin{equation} \rightarrow ~ ~ ~
\frac{\text{d}\ec_{04}} {\text{d} z} =
 -\frac{i \omega_{4}}{2 c}
\frac{2 \hbar e^{4} K}{360\pi^{2} m^{4} c^{7} \epsilon_{0}}
\ec_{01} \ec_{02}\ec_{03} \label{eqref}
\end{equation}
$K$ is a factor that depends on the directions of the incident
beams and of their polarisation ($K<14$). In our configuration, we
have $K\approx 0.56$\cite{bientot}.

The equations describing the growth rate of $\ec_{04}$ of 4 wave
mixing in a low pressure gas (eq. \ref{danslegaz}) and of stimulated
photon scattering in vacuum (eq. \ref{eqref}) have the same
form.
Therefore we can define\cite{bientot} the QED susceptibility of vacuum~:
\begin{equation}
\chi^{(3)}_{v}=
\frac{2 \hbar e^{4} K}{360\pi^{2} m^{4} c^{7} \epsilon_{0}}
=\frac{K}{45\pi\alpha} \left(\frac{r_{e} e}{m c^{2}}\right)^{2}
\approx 3.0~10^{-41}K~(\text{m}^{2}/\text{V}^{2})
\end{equation}

\section{Experimental results}

\subsection{Four wave mixing in a gas}
The typical sensitivity of the delay of the third beam is on the order of
20~fs, which shows that the duration of the beam provided by the OPA 
is similar to that of the pump beam.

The main source of fluctuation of the $\chi^{(3)}$ signal is caused by a vertical
oscillation of beam \#2 due to the pumping system of the compressor.
This produces a periodic variation of the signal
with an average loss factor equal to 5.
The search of a signal in vacuum was interspersed by the observation
of the $\chi^{(3)}$ signal in the gas jet.
This signal showed an excellent long term
stability~: after a fraction of an hour the $\chi^{(3)}$ signal was
still present, and of the same order of magnitude.
The $\chi^{(3)}$ signal was observed with laser energies set just
below plasma threshold.
We can obtain an upper bound of this laser energy from the intensity
threshold $I_{1}$ of ionization of nitrogen, close to
$10^{14}\text{W}/\text{cm}^{2}$~:
\begin{equation}
E = \frac{\pi^{3/2}}{4\sqrt{\ln{2}}} \tau I_{1} w^{2}
\end{equation}
that is $E\approx 1~\mu$J. Up to $5.~10^{4}$ photons were observed.

\begin{figure}[hbt]
\begin{center}
 \mbox{\epsfig{figure=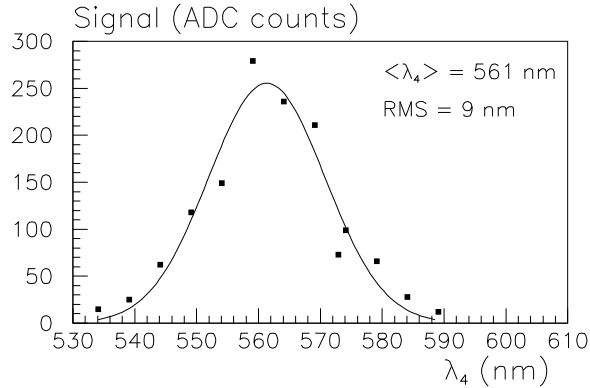,width=0.48\linewidth}}
\caption{Spectrum of the $\chi^{(3)}$ signal, with a gaussian fit.
The maximal value in 200 shots is shown as a function of the central
value of the spectrometer, with an exit slit of 3~mm, 
equivalent to a spectral range of 9~nm.}
 \label{spectrum}
 \end{center}
\end{figure}

The origin of the signal is identified as four wave mixing in the gas,
because it is present only with the three beams injected.
Furthermore, its spectrum is found to peak at the wavelength
$\lambda_{4}$ of four wave mixing (fig. \ref{spectrum}).

The FWHM spectral width, $\Delta\lambda_{4}=22~\text{nm}$, corresponds
to a Fourier limited FWHM duration of $22~\text{fs}$.
To our knowledge, this is the first observation of large angle four
wave mixing.

We compute the expected number of observed scattered photons from the
integration of equation (\ref{danslegaz}). We get approximately~:
\begin{equation}
N_{4,\text{N}_{2}}= 
\epsilon_{PM} \cdot
\epsilon_{Sp} \cdot
\epsilon_{Osc} \cdot
\frac{128}{\pi\sqrt{3}^{3}}
\frac{(\hbar \omega_{4}) E_{1} E_{2} E_{3}}{e^{4}w^{2} (c\tau)^{2} }
(\chi^{(3)}_{\text{N}_{2}})^{2}
\end{equation}
where $E_{i}$ are the energy of the three incoming laser pulses, 
$\epsilon_{PM}$, 
$\epsilon_{Sp}$, 
$\epsilon_{Osc}$, 
are the quantum efficiency of the PMT, the transmission of the
spectrometer, and the loss factor due to a transverse oscillation of
beam 2.

The third order susceptibility of nitrogen has been measured by other
experiments, in different configurations.
Nibbering {\em et al.} have measured the red shift of short pulse
spectra due to self phase modulation (SPM) \cite{georges}.
The value of the nonlinear refractive index, measured at 1 bar, is
$\displaystyle n_{2}=
2.3~10^{-23}\text{m}^{2}/\text{W}$\cite{georges}.
That value of $n_{2}$ is related to the third order susceptibility by~:
$\displaystyle n_{2}=\chi^{(3)}_{\text{N}_{2},\text{SPM}}/(c\epsilon_{0})$,
 so that 
$\chi^{(3)}_{\text{N}_{2},\text{SPM}} \approx 6.1~10^{-26}
\text{m}^{2}/\text{V}^{2}$. 

Note that the ratio of the third order susceptibilities in vacuum and
in gas with pressure $P_{\text{bar}}$ is
$\chi^{(3)}_{v}/\chi^{(3)}_{\text{N}_{2},\text{SPM}}
\approx 4.8~10^{-16} \times K/P_{\text{bar}}$ :
the QED vacuum is indeed linear to a very good approximation.
The two contributions (four wave mixing in a gas and QED stimulated
photon scattering) are of the same order of magnitude only for a
pressure close to 
$P_{\text{Lim}}\approx 4.8~10^{-13}\times K~\text{mbar}$.

Here, at a pressure of about 0.3~bar, the expected number of photons is
$N_{4,\text{N}_{2}}=3.5~10^{6}$.

The value of $\chi^{(3)}_{\text{N}_{2}}$ has been also measured by
Lehmeier {\em et al.} in third harmonic generation (THG) in Nitrogen
by a picosecond Nd:glass laser pulse\cite{thg}.
The obtained value, $\chi^{(3)}_{\text{N}_{2}\text{THG}}=
6.7~10^{-27}\text{m}^{2}/\text{V}^{2}$, is about ten times lower than
for self phase modulation, and the corresponding value of
$N_{4,\text{N}_{2}}$ is $4.2~10^{4}$.

These numbers are of the same order of magnitude of the number of
photons observed in this experiment.

\subsection{Stimulated Photon scattering in vacuum}
A signal was searched for in vacuum with laser energies of 150~mJ,
55~mJ and 200~$\mu$J, and with a spectral acceptance of 30~nm.
After compression, transport, and taking into account
only  the energy that is contained in the central spot at focus,
only a fraction of the laser energy is actually available.
An estimate of that fraction has been obtained by a subsequent
experiment, that has studied precisely the threshold of Helium
ionization by a single beam\cite{rapha}.
 We use here a conservative number of 3\%.

\begin{figure}[hbt]
\begin{center}
 \mbox{\epsfig{figure=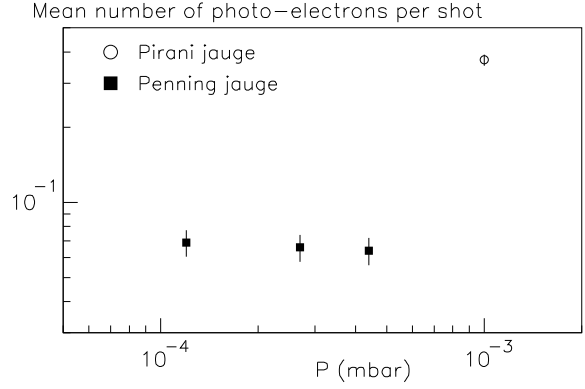,width=0.47\linewidth}}
\caption{Pressure dependence of the noise.}
 \label{pressure}
 \end{center}
 \end{figure}

\begin{figure}[hbt]
\begin{center}
 \mbox{\epsfig{figure=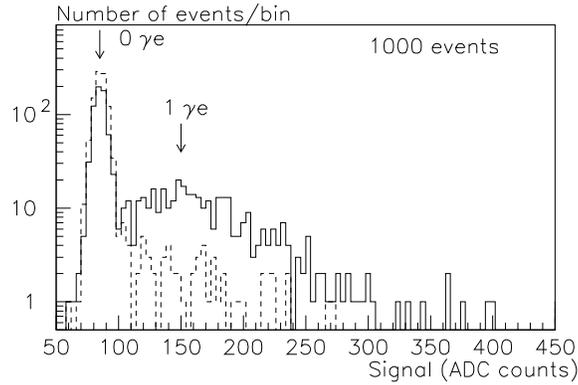,width=0.47\linewidth}}
\caption{Spectrum of the signal for pressure $P=10^{-3}\text{mbar}$ (solid line), and 
 $P=1.7~10^{-4}\text{mbar}$ (dashed line). The signal from the PMT
 was amplified with gain 80 before digitization. The one
 photo-electron spectrum is clearly visible over the pedestal (no
 photo-electron).} 
 \label{spectreADC}
 \end{center}
\end{figure}

At high residual pressure ($P> 5.~10^{-4}\text{mbar}$), we observe a
BG noise from the residual plasma (fig. \ref{pressure}).
The one photo-electron signal is easily identified from the pedestal
(fig. \ref{spectreADC}).

At lower pressure, the remaining noise is due to the creation of white
light on a specific imperfect part of the optics that could not be
fixed.

Data were taken at $10^{-4}~\text{mbar}$, with an integration time of
100~s; that is 1000 laser shots.
For most of the laser shots, no photo-electron ($\gamma_{e^{-}}$) was
detected in the PMT. The number of laser shots, with at least 
$1 \gamma_{e^{-}}$ is presented in the following table. 
\begin{center}\begin{tabular} {lr}
 beam 1 alone & 19 $\gamma_{e^{-}}$\\
 beam 2 alone & 42 $\gamma_{e^{-}}$\\
 beam 3 alone &  5 $\gamma_{e^{-}}$\\ \hline
 total        & 66 $\gamma_{e^{-}}$\\
\\
3 beams together & 60 $\gamma_{e^{-}}$\\
\end{tabular}\end{center}
No evidence for an excess non linear contribution of the three beams
was observed.

\section{Experimental limit of the elastic cross section}

We derive an upper bound of the elastic cross section by the use of a
given model -- ``chosen to be'' here QED~:
\begin{equation}
\sigma_{\text{Lim}} = \frac{N_{4,\text{obs}}}{N_{4,\text{QED}}}
\times \sigma_{\text{QED}}
\end{equation}

We compute the expected number of scattered photons $N_{4,\text{QED}}$
from the integration of equation (\ref{eqref}). We get approximately~:
\begin{equation}
N_{4,\text{QED}}= 
\epsilon_{PM} \cdot
\epsilon_{Sp} \cdot
\epsilon_{Osc} \cdot
\frac{16}{2025} 
\left(\frac{2}{\pi\sqrt{3}}\right)^{3}
\frac{(\hbar \omega_{4}) E_{1} E_{2} E_{3}}{(m c^2)^{4}}
\frac{r_{e}^{4}}{w^{2} (c\tau)^{2}}
K^{2}
\end{equation}
where $E_{i}$ are the energy of the three incoming laser pulses, 
$\epsilon_{PM}$, 
$\epsilon_{Sp}$, 
$\epsilon_{Osc}$, 
are the quantum efficiency of the PMT, the transmission of the
spectrometer, and the loss factor due to a transverse oscillation of
beam 2.

\begin{figure}[htb] \begin{center}
\mbox{\epsfig{figure=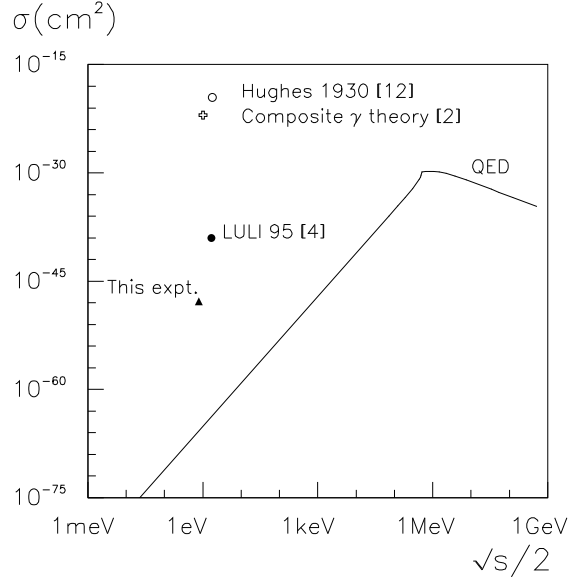,width=0.47\linewidth}}
\caption{Elastic photon cross section as a function of photon cms energy.
\label{sectioneff}}
\end{center} \end{figure}

We obtain finally a QED prediction of $N_{4,\text{QED}}\approx
7\cdot 10^{-21}$ per shot while the observed limit is
$N_{4,\text{obs}}\approx 6\cdot 10^{-3}$ per shot.
The elastic QED cross-section at $\hbar \omega =0.8~\text{eV}$ is 
$\sigma_{\text{QED}}=1.8~10^{-66}~\text{cm}^{2}$.
The obtained limit is therefore
$\sigma_{\text{Lim}}=1.5~10^{-48}\text{cm}^{2}$, that is 18
orders of magnitude from QED (fig. \ref{sectioneff}).

\section{Conclusion}

We have searched for stimulated photon scattering at a cms photon
energy of 0.8~eV.
The spatial and temporal overlap of three $4~\mu$m, 40~fs laser beams
has been obtained.
The last step in the alignment procedure is the maximisation of four
wave mixing in a gas in exactly the same configuration as for
$\gamma\gamma$ scattering.
To our knowledge, this is the first observation of large angle 4 wave
mixing in a gas.

In vacuum, no evidence for $\gamma\gamma$ scattering was observed.
We obtain an approximate improved upper limit of the cross section of
$\sigma_{\text{Lim}}=1.5~10^{-48}\text{cm}^{2}$, at 18 orders of
magnitude from QED.
This is an improvement of nine orders of magnitude compared to the
previous, non stimulated experiment\cite{z}. 

Several orders of magnitude could be gained by an improvement in the
operation of the laser, the OPA, by an increase of the available fraction
of the laser energy in the central spot at focus, by further work on
the background noise, and by fixing the transverse oscillation of one
beam.

The actual observation of the QED effect will wait for the availability of
short pulse lasers in the 10~J class, probably in the next decade.

~

This work has been funded by Training and Mobility of Researchers
contracts \#~ERBFMGECT950019.

\end{document}